\documentclass[manuscript]{aastex}

\usepackage{aastexug}   

\slugcomment{Accepted by The Astronomical Journal on August 28, 2002}

\shortauthors{Aller \& Richstone}
\shorttitle{Black Hole Mass Density}

\begin{document}
\newcommand{\tidl}{\tilde{L}}
\newcommand{\tidsigma}{\tilde{\sigma}}
\newcommand{\tidmbh}{\tilde{{\cal{M}}_\bullet}}

\title{The Cosmic Density of Massive Black Holes from Galaxy Velocity Dispersions}

\author{M.C. Aller\altaffilmark{1} and  D. Richstone\altaffilmark{2}}
\affil{Astronomy Department, University of Michigan, 
	Ann Arbor, MI 48109}

\altaffiltext{1}{maller@astro.lsa.umich.edu}
\altaffiltext{2}{dor@astro.lsa.umich.edu}


\begin{abstract}

Supermassive black holes are thought to be relics of 
quasars, and their numbers and masses are therefore related to the
quasar luminosity function and its evolution with redshift.  
We have used the relationship between black hole mass and bulge
velocity dispersion (the $M_\bullet - \sigma$ relation) to make an
improved estimate of the mass density and mass spectrum of supermassive black holes.
Uncertainties in the $M_\bullet - \sigma$ relation have little effect
on the mass density. We find a mass density of
${(4.8 \pm 1.6) h^2 \cdot 10^5} {{M_{\odot}} {{Mpc}^{-3}}}$. 
Some of the variance in published density 
estimates comes from the use of different values of the Hubble
constant.  

\end{abstract}


\keywords{galaxies: general, nuclei, black holes}

%

\section{Introduction}

It is now fairly well established that most or all bulges contain
supermassive black holes \citep{Richstone,Kormendy,Magorrian}.
Within the last two years it has become
clear that the mass of the black hole (BH) is related to the velocity
dispersion of the bulge of the host galaxy.  This relationship has
less scatter than a previously noted relationship between the black
hole mass and the bulge light \citep{Gebhardt,Merritt,Kormendy}. 
It has also been known
for some time that the masses, and cosmic mass density of BHs, crudely
correspond to the masses and numbers predicted by the luminous output
of quasars, under the assumption that they are powered by accreting
black holes with radiative efficiencies near 10\%.  Published
estimates of the BH mass density predicted by quasars are around $2
\cdot 10^5 {M_{\odot}} {{Mpc}^{-3}}$ \citep{Chokshi, Soltan, Salucci}, while estimates
from the X-ray background tend to be larger.  Using the X-ray background, 
\citet{Salucci} obtain an estimate of 
$3-5 \cdot 10^5 {M_{\odot}} {{Mpc}^{-3}}$, 
while \citet{Fabian} estimate the mass density to be 
$6-9 \cdot 10^5 {M_{\odot}} {{Mpc}^{-3}}$. 

It is useful to note that the above density estimates are independent
of the Hubble constant or cosmological model (so long as it is general
relativistic, isotropic and homogeneous) because they can be cast in
terms of the surface brightness of quasar light on the sky, and hence
determine its local energy density.  Since the mass density of the radiating
objects declines as the universe expands as $(1+z)^3$ and the photon
energy of their radiation declines as $(1+z)^4$, the only cosmological
correction connecting the observed radiation to the local BH density
is $(1+z)$, where z is the redshift of the epoch of the emission of the radiation.  

On the other hand, the local density of supermassive BHs are obtained
by multiplying the density of galaxies by the typical BH mass per
galaxy. \citet{KR} used the BH 
mass-bulge luminosity relationship, together with the \citet{Loveday} 
estimate of the luminosity density in galaxies, to estimate the local
density of BHs; for a Hubble constant of 80 
km/s/Mpc this is $1 \cdot 10^6 {M_{\odot}} {{Mpc}^{-3}}$.
More recently, \citet{MF} estimated this same quantity using
the newer $M - \sigma$ relation.  Their approach was to use
the $M - \sigma$ relation to calibrate a mass-luminosity relation and
then multiply by the galaxy luminosity function. 

In this paper we take an alternate approach.  We use multiple
estimates of the luminosity functions of galaxies for different Hubble
types, together with estimates of bulge luminosity as a function of
Hubble type, together with observed relationships between luminosity
and velocity dispersions or rotation velocity, to construct a
'velocity function' (the number of galaxies per magnitude per unit
volume) for bulges as a function of Hubble type.  This function can
then be used together with the $M-\sigma$ relation to estimate BH mass
functions.  Our basic result using this procedure is that the mean
mass density of supermassive BHs is $4.8 {h_{100}}^{2}\cdot 10^5 {M_{\odot}} {{Mpc}^{-3}}$.

After we completed much of the analysis in this paper, we became aware
of an estimate by \citet{Yu} based on the Sloan Digital Sky Survey (SDSS), that
gives an estimate of 
$\rho_{\bullet} = (2.5 \pm 0.4) {h_{0.65}}^2 \cdot 10^5 {M_{\odot}} {{Mpc}^{-3}}$.
By assuming that h=0.65, as in their paper, our results are comparable with 
theirs.
The rather surprising conclusions about the radiative efficiency of
quasars derived by \citet{Yu} from their low value of
$\rho_{\bullet}$ illustrate the importance of an improved estimate of this
parameter.

\section{Method}

The basic assumption underlying this note is that the bulge velocity
dispersion of a galaxy can be used to predict the mass of its
supermassive black hole.  Under that assumption, the best way to
proceed would be to directly observe velocity dispersions of a
volume--limited sample of galaxy bulges and to integrate over the
implied distribution.  Such a catalogue might well exist in the
future, 
\footnote{After starting this project, the early release data for the
SDSS became available, but we pursued this calculation from completely
independent data.}
but at present we adopt a simpler expediency. We use a power law
\citep{fj} to convert various luminosity functions to a ``dispersion
function'', and the $M-\sigma$ relation for black holes to
convert the dispersion function to a BH mass function.

In general, if there are $N$ galaxies per cubic megaparsec brighter
than a luminosity $L$, then the luminosity function is 
\begin{equation}
{dN \over dL} \equiv \Phi(L). 
\end{equation}
This can be directly converted to the BH mass function by using the
derivative of the relationship between luminosity and the velocity
dispersion ($\sigma$) and the black hole mass (${\cal{M}}_\bullet$) and $\sigma$ for
bulges. Without any loss of generality, it can be written as
\begin{equation}
\Theta({\cal{M}}_\bullet) \equiv {dN \over d{\cal{M}}_\bullet} =
 \Phi(L) \, \left({dL \over
   d\sigma}\right) \left( { d{\cal{M}}_\bullet \over d\sigma} \right)^{-1}
\end{equation}

Since in what follows we will rely on luminosity
functions due to \citet{Marzke}, we adopt his parameterization of the
luminosity function of different types of galaxies as Schechter
functions and adopt power laws for the two relationships.  
To simplify the algebra we normalize luminosity, dispersion and black
hole mass by their values at the fiducial luminosity of the Schechter
function $L_*$, as follows 
\begin{eqnarray}
	\tidl &= L/L_* \\
	\tidsigma &= \sigma/\sigma_* \\
	\tidmbh &= {\cal{M}}_\bullet / {\cal{M}}_{\bullet*}
\end{eqnarray}

We then can write the Schechter parameterization and the relationships
between luminosity and dispersion and dispersion and ${\cal{M}}_\bullet$ as
\begin{eqnarray}
{dN \over dL} &= \Phi(L) = \left({\phi_* \over L_*}\right)
        \tidl^\alpha 
	e^{-\tidl} \\
 ~ & \tidl = \tidsigma^n \\  
 ~ & \tidmbh = \tidsigma^\lambda   
\end{eqnarray}
with $\phi_*$ in $Mpc^{-3}$.
Substituting these equations into equation 2 gives the following
dispersion function
\begin{equation}
{dN \over d\sigma} = \left({ n \phi_* \over \sigma_* }\right)
        \tidsigma^{\beta-1}
	e^{-\tidsigma^n} 
\end{equation}
with $\beta = n(\alpha + 1)$ and the following mass 
function for supermassive black holes 
\begin{equation}
{dN \over d{\cal{M}}_\bullet} = \Theta({\cal{M}}_\bullet) = 
      \theta_* {{\cal{M}}_{\bullet*}}^{-1} \tidmbh^{\gamma - 1}\,
	e^{-\tidmbh^\epsilon} 
\end{equation}
with $\theta_* = \epsilon \phi_*$, $\epsilon = n/\lambda$
and $\gamma = \beta/\lambda = n(\alpha + 1) / \lambda$. 

Unless explicitly parametrized by 
h = $ H_o / (100 km \cdot s^{-1} \cdot Mpc^{-1})$, all calculations were done assuming 
$ {H_o} = 100 km \cdot s^{-1} \cdot Mpc^{-1}$.

%
%

\section{Results} 

In order to determine the number density as a function of dispersion,
the Schechter function fits, valid for ${M_Z} \geq -21.5$
(corresponding to $\sigma_* = 400$ km/s and 
${{\cal{M}_{{\bullet}*}}} = 1.8 \cdot 10^9 M_{\odot}$ for an elliptical
galaxy), obtained by
 \citet{Marzke}, were
modified using the corresponding bulge to total galaxy luminosity ratios,
calculated from the data given in \citet{Simien} for the appropriate galaxy
range, as
given in Table 1, to produce a fit as a function
of the bulge-only luminosity for the spiral galaxies: 
\begin{eqnarray}
{dN \over dL_{bulge}} &= \Phi(L_{bulge}) = \left({h^3 \cdot \phi_* \over L_{*bulge}}\right)
        \tidl_{bulge}^\alpha
        e^{-\tidl_{bulge}} \\
 ~ & \tidl_{bulge} = L_{bulge}/L_{*{bulge}} 
\end{eqnarray}
A power-law relation of the form
\begin{equation}
{\tidl = \tidsigma^n \\}
\end{equation}
where $\tidl$ refers to the total luminosity ratio for the elliptical
galaxies and to the bulge-only luminosity ratio for the spiral galaxies,
may be used to convert the function of luminosity (6, 11) to one of dispersion
such that
\begin{equation}
{dN \over d\sigma} = h^3 \cdot \left({ n \phi_* \over \sigma_* }\right)
        \tidsigma^{\beta-1}
        e^{-\tidsigma^n} \\
\end{equation}
with $\beta = n(\alpha + 1)$. 

To obtain a value for the parameter $\sigma_*$ for each galaxy type, the
parameter $M_*$ from the \citet{Marzke} Schechter function fits 
are converted to the equivalent dispersion,  $\sigma_*$ 
using a relationship 
between the magnitude and the dispersion given by \citet{deV}. The "best"
relationship between magnitude and dispersion, selected from several
in this paper, was determined by  
\citet{Gonzalez} to be:
\begin{equation}
{-(M_{B_T} + 5log(h))} = a + b(log(\sigma)-2.3) 
\end{equation}
with $a = 19.71 \pm 0.08$ and $b = 7.7 \pm 0.7$.
This results in a slope for
the power-law relation (13) of $n = b/2.5 = 3.08 \pm 0.28$.
However, this equation is in the $M_{B_T}$ magnitude system, and so must be
converted into the Zwicky magnitude system in order to be combined
with the \citet{Marzke} Schechter function fitting parameters. Despite concerns
that there may not be a linear relationship between these two magnitude systems
\citep{MHG}, an approximate conversion was adopted by combining the conversion given
in \citet{Shanks} 
\begin{equation}
{M_{b_j}} = {M_Z} - 0.45
\end{equation}
with that given in \citet{Gonzalez}
\begin{equation}
{M_{b_j}} \approx {M_{B_T}} + 0.06 
\end{equation}
to obtain
\begin{equation}
{M_{B_T}} \approx {M_Z} - 0.51
\end{equation}
which allows the magnitude-dispersion relation to be written as
\begin{equation}
{-M_Z} + 0.51 = a + b(log(\sigma)-2.3) + 5log(h)
\end{equation}
This relation may then be combined with the \citet{Simien}
magnitude correction for the bulge-to-total
galaxy magnitude ratio, $\Delta M$,
so that only the ${M_{*bulge}}$ is used to calculate
the dispersion for the spiral galaxies, such that 
\begin{equation}
-({M_Z} - 0.51 + {\Delta}M) = a + 
b(log(\sigma)-2.3) + 5log(h)
\end{equation}
Hence, the velocity dispersion, ${\sigma_*}$, may be expressed as
\begin{equation}
{\sigma_*} = 199.5 \cdot 10^{\chi}\\
\end{equation}
with
\begin{equation}
\chi = ({{{-M_{Z*}}+0.51- a - 5log(h)-{\Delta}M})b^{-1}}
\end{equation}

The number density as a function of dispersion (14) may then be combined with
the relation between the dispersion and the black hole mass given by
\citet{Tremaine}, corrected from h=0.8 (to h=1) for consistency, 
\begin{equation}
{{\cal{M}_{\bullet}}} = h^{-1} \cdot 0.8 \cdot 10^{8.13 \pm 0.06} \cdot \left({{\sigma} 
\over 200}\right)^{\lambda}
\end{equation} 
(with $\lambda = 4.02 \pm 0.32$)
to obtain the final equation for the number density as a function
of black hole mass:
\begin{equation}
{dN \over d{\cal{M}}_\bullet} = \Theta({\cal{M}}_\bullet) =
      h^{3} \cdot \theta_* {{\cal{M}}_{\bullet*}}^{-1} \tidmbh^{\gamma - 1}\,
        e^{-\tidmbh^\epsilon}
\end{equation}
again with $\theta_* = \epsilon \phi_*$, $\epsilon = n/\lambda = 0.8 \pm 
0.1$,
$\gamma = \beta/\lambda$ and
\begin{equation}
{{\cal{M}}_{\bullet*}} = h^{-1} \cdot 0.8 \cdot 10^{8.13 \pm 0.06} \cdot 10^{\lambda \cdot 
\chi}
\end{equation}
This method of calculating ${\cal{M}}_{\bullet*}$, rather than directly
applying a relationship between ${\cal{M}}_{\bullet} - L_{bulge}$, was 
used because \citet{Gebhardt} indicate that there is a tighter correlation
between ${\cal{M}}_{\bullet} - \sigma$ than between 
${\cal{M}}_{\bullet} - L_{bulge}$.

Using expression (24) for the number density, an
expression for the total mass density of black holes may be obtained by integrating
the function
\begin{eqnarray}
\rho &  = \int_{{\cal{M}}_{\bullet{min}}}^{ \infty} {\cal{M}}_{\bullet}
 \Theta({\cal{M}}_\bullet)\, d{\cal{M}}_{\bullet}\\
~ &   = h^{2} \cdot \int_{{\cal{M}}_{\bullet{min}}}^{\infty} \theta_* \tidmbh^{\gamma}\,
        e^{-\tidmbh^\epsilon} \, d{\cal{M}}_{\bullet}  
\end{eqnarray}
from an assumed minimum black hole mass to an infinite black hole mass 
for each galaxy type, and then summing the results to obtain
an estimate for the total mass density of black holes. The calculation
was done adopting a minimum black hole mass limit of ${10^6} M_{\odot}$, 
which results in a total mass density for all galaxy types of
\begin{equation}
\rho_{\bullet} = {(4.8 \pm 1.6) h^2 \cdot 10^5} {{M_{\odot}} {{Mpc}^{-3}}} 
\end{equation}
The results of this calculation are given in Table 2; along with the 
number density function fitting parameters for each galaxy type. The
uncertainties and average values given for each of these parameters
were derived using a Monte Carlo analysis, with the largest source of uncertainty
stemming from the bulge-to-total galaxy luminosity ratios given in Table 1. Figures
1-4 show histograms of the resulting densities from the Monte Carlo calculation,
which illustrate that the median value for the density tends to be slightly lower
than the average value, but still well within the
stated uncertainties of the density, with a relatively symmetric distribution of
densities.

The calculation of the mass density was then redone using exactly the same
method, but utilizing completely different
galaxy luminosity functions obtained from \citet{Madgwick}. 
In \citet{Madgwick}
the galaxies were divided into subsets based on spectral type, $\eta$, rather
than by Hubble T-type. The parameter $\eta$ is determined based on the 
absorption and emission line strengths in the galaxy spectrum. In order to 
apply the method developed in this paper, it was necessary to convert the
spectral types into T-types so that the appropriate bulge-to-total luminosity
conversions could be applied. This was done by utilizing the data given
in Figure 4 of \citet {Madgwick} which showed the correlation between
$\eta$ and morphological type, and assuming a relatively even distribution
of $\eta$ over T-types. For the luminosity function designated as Type 1
($\eta < -1.4$), an assumption was made that galaxies were evenly divided
between elliptical, S0, and Sa galaxies for the purposes of this calculation. 

The resulting mass density obtained using the \citet{Madgwick} luminosity
functions is
\begin{equation}
\rho_{\bullet} = {(6.9 \pm 1.4) h^2 \cdot 10^5} {{M_{\odot}} {{Mpc}^{-3}}}
\end{equation}
with the results for the individual spectral types 
 given in Table 3.
This is slightly larger than the result obtained using the Marzke
functions; however, it is strongly dependent on the assumptions made
regarding the division into morphological type, particularly of the
\citet {Madgwick} Type 1 galaxies. If is is assumed that all of
the Type 1 galaxies are elliptical, the resulting mass density
becomes ${(10.5 \pm 2.0) h^2 \cdot 10^5} {{M_{\odot}} {{Mpc}^{-3}}}$, if
it is assumed that they are all S0 galaxies then the mass density
becomes ${(6.0 \pm 2.0) h^2 \cdot 10^5} {{M_{\odot}} {{Mpc}^{-3}}}$,
and if it is assumed that they are all Sa galaxies, then the
mass density is calculated to be
${(4.0 \pm 1.8) h^2 \cdot 10^5} {{M_{\odot}} {{Mpc}^{-3}}}$. Considering 
the uncertainties in using the \citet{Madgwick} spectral
type classifications for the purpose
of applying a bulge-to-disk luminosity correction, these results
are consistent with those obtained using \citet{Marzke}. Also of importance
is the fact that \citet{Madgwick} specifically take into account issues
of Malmquist bias when determining their luminosity functions. 
\citet{Marzke} make no mention of this. Therefore, it appears that
this possible omission will not have a significant impact on the final
density of black holes.

\section{Discussion}
 
In Table 2, 
as expected, the ellipticals have the
largest central black hole mass, ${\cal{M}}_{\bullet*}$, values, while later-type galaxies have
increasingly smaller ${\cal{M}}_{\bullet*}$ values, corresponding to their
smaller relative bulge sizes. Plots of the number function (Figure 5) and
the mass function (Figure 6) illustrate that the contribution from the S0
galaxies dominates at low ${\cal{M}}_{\bullet}$, while the ellipticals
dominate for high ${\cal{M}}_{\bullet}$ values. Overall, the strongest
contribution to the total mass density is from S0 galaxies, with important
contributions from the elliptical and Sa-Sb galaxies, but a small 
contribution from the Sc-Sd galaxies.  

The total cosmic mass density of black holes of 
${(4.8 \pm 1.6) h^2 \cdot 10^5} {{M_{\odot}} {{Mpc}^{-3}}}$ obtained
from this calculation is relatively independent of several of the
adopted parameters. As illustrated by Figure 7, decreasing the 
lower limit of integration in the calculation of the mass density
does not significantly change the resulting mass density estimate. This
is because the black holes located in small-bulge galaxies are, themselves,
small, and there are not enough small-bulge galaxies to make up in number
for this lack in mass. This is further illustrated by the flat slope
in the plot of the mass function (Figure 6). 
Similarly, altering the value of $\lambda$, the power law index,
in the relationship between the black hole mass, ${\cal{M}}_{\bullet}$, and the 
velocity dispersion, $\sigma$ (eqns 5 \& 8), has a small effect 
on the
final density. This is because most of the contribution is in the range of
$10^7 - 10^8 M_{\odot}$ - the same region in which the ${\cal{M}}_{{\bullet}*}$ for
this relation is
located, and in which the relationship is determined best.  
A perturbation in the power law index for masses near
this mass will have little effect on the final answer. Changing the
zero-point, however, would have a roughly linear effect on the resulting mass
density, although for the currently accepted relationships, the differences
in the zero-point are small. 
 Redoing the calculation using the best-estimate \citet{Merritt2} relation stating that
\begin{equation}
{{\cal{M}_{\bullet}}} = (1.48 \pm 0.24) \cdot 10^{8} \cdot \left({{\sigma}
\over 200}\right)^{(4.65 \pm 0.48)}
\end{equation}
instead of the \citet{Tremaine}
relation (23),
produces a result of $(4.3 \pm 1.6) h^2 \cdot 10^5 {{M_{\odot}} {{Mpc}^{-3}}}$, which
agrees well with our previous estimate. This estimate for the mass density also
agrees, within the limits of the uncertainty, with that given in the \citet{Merritt2}
paper of $\rho_{\bullet} \approx 3 \cdot 10^5 {{M_{\odot}} {{Mpc}^{-3}}}$.

The parameters with the most significant impact on the resulting
density, other than the $M-\sigma$ relationship from \citet{deV} and the
Schechter functions from \citet{Marzke}, are the assumptions about
the morphological distributions of galaxies and the resulting
bulge-to-disk luminosity corrections, adopted from \citet{Simien}.
This is illustrated by the results obtained using the galaxy luminosity
functions from \citet{Madgwick} instead of \citet{Marzke}. By assigning
a different morphological distribution to the \citet{Madgwick} Type 1
luminosity function galaxies, the resulting mass density varies by a
factor of approximately 2.5. Given this uncertainty in the
mass density resulting from the calculation using the \citet{Madgwick} data,
we are more confident in our estimates derived from the \citet{Marzke}
luminosity functions. 

The Schechter function fits may be less applicable in the low-black-hole mass
regime, particularly for ${\cal{M}}_{\bullet} < 10^6 M_{\odot}$, and to a lesser
degree for galaxies with ${\cal{M}}_{\bullet} < 10^7 M_{\odot}$. While there is information
on black holes down to $10^6 M_{\odot}$, in our own galaxy and M32, the Schechter
function fits may not correctly apply to elliptical galaxies, may include
two types of galaxies within one classification, or may mix exponential and
de Vaucouleurs-law galaxies together.
   We assume that while the Schechter function
fits for elliptical galaxies may be dominated by dE galaxies at low luminosities,
the same ${\cal{M}}_{\bullet} - \sigma$ relationship is still applicable;
in our own galaxy and a small number of other galaxies with
exponential profiles and measured black hole masses, this relationship appears
to hold true.

It must be noted that the resulting mass density of black holes may be an underestimate
due to a possible bias in the relationship between the black hole
mass and the galaxy velocity dispersion (pointed out to us by Dr. Scott Tremaine). This bias
relates to the fact that for any given velocity dispersion, there is a range of
black hole masses, and this distribution may not be symmetrical about the ridgeline.
We have investigated this possible bias by assuming a log-normal 
distribution of black hole masses, with a disperson of 0.15 in the log space, and recomputing the
black hole mass density. This results in a mass density of 
$\rho_{\bullet} = {(5.0 \pm 2.1) h^2 \cdot 10^5} {{M_{\odot}} {{Mpc}^{-3}}}$, which
is 4 \% larger than our original estimate.

It is often useful to have a simple approximation available for crude
calculations. While the number functions we have derived are not of the 
Schechter form, we can find approximating functions of the form
\begin{equation}
{dN \over d{\cal{M}}_\bullet} =
c \cdot \tidmbh^{-\alpha} e^{-\tidmbh}
\end{equation}
For the number function based on the \citet{Marzke} data, the roll off at high
mass is too gradual to be well-approximated by any function of the form (31). 
We illustrate a fair fit in Figure 8. This fit, which works best near
$\cal{M}_{\bullet} \approx$ $10^8 M_\odot$, where the information on the numbers
of black holes is the most extensive, has the parameters
(c, $\cal{M}_{\bullet*}$, $\alpha$) $= (1.3 \cdot 10^{-10} {M_\odot}^{-1}Mpc^{-3},
7.9 \cdot 10^7 M_\odot, 0.95)$. We can find a better fit to the number function
produced by the \citet{Madgwick} luminosity function. It is shown in Figure 9. 
The parameters for this fit are 
(c, $\cal{M}_{\bullet*}$, $\alpha$) $= (3.2 \cdot 10^{-11} {M_\odot}^{-1}Mpc^{-3},
1.3 \cdot 10^8 M_\odot, 1.25)$.   

The recent paper by \citet{Yu} utilizes luminosity and velocity dispersion functions 
from early-type galaxies obtained as part of the Sloan Digital Sky Survey to 
perform a similar calculation. They obtain a result of 
$(2.5 \pm 0.4) {h_{0.65}}^2 \cdot 10^5 {M_{\odot}} {{Mpc}^{-3}}$. 
If we take h=0.65 to compare our result to \citet{Yu},
we obtain 
${\rho_{\bullet} = (2.0 \pm 0.7) \cdot 10^5} {{M_{\odot}} {{Mpc}^{-3}}}$, 
which is in reasonable agreement with their result, particularly when the possible 4 \%
bias is taken into account.

We acknowledge support for proposals 7388, 8591, 9106 and 9105,
provided by NASA through grants from the Space Telescope Science
Institute, which is operated by the Association of Universities for
Research in Astronomy, Inc., under NASA contract NAS 5-265555.

%
%
%
%
%
%
%

{}

\clearpage

\begin{figure}
\plotone{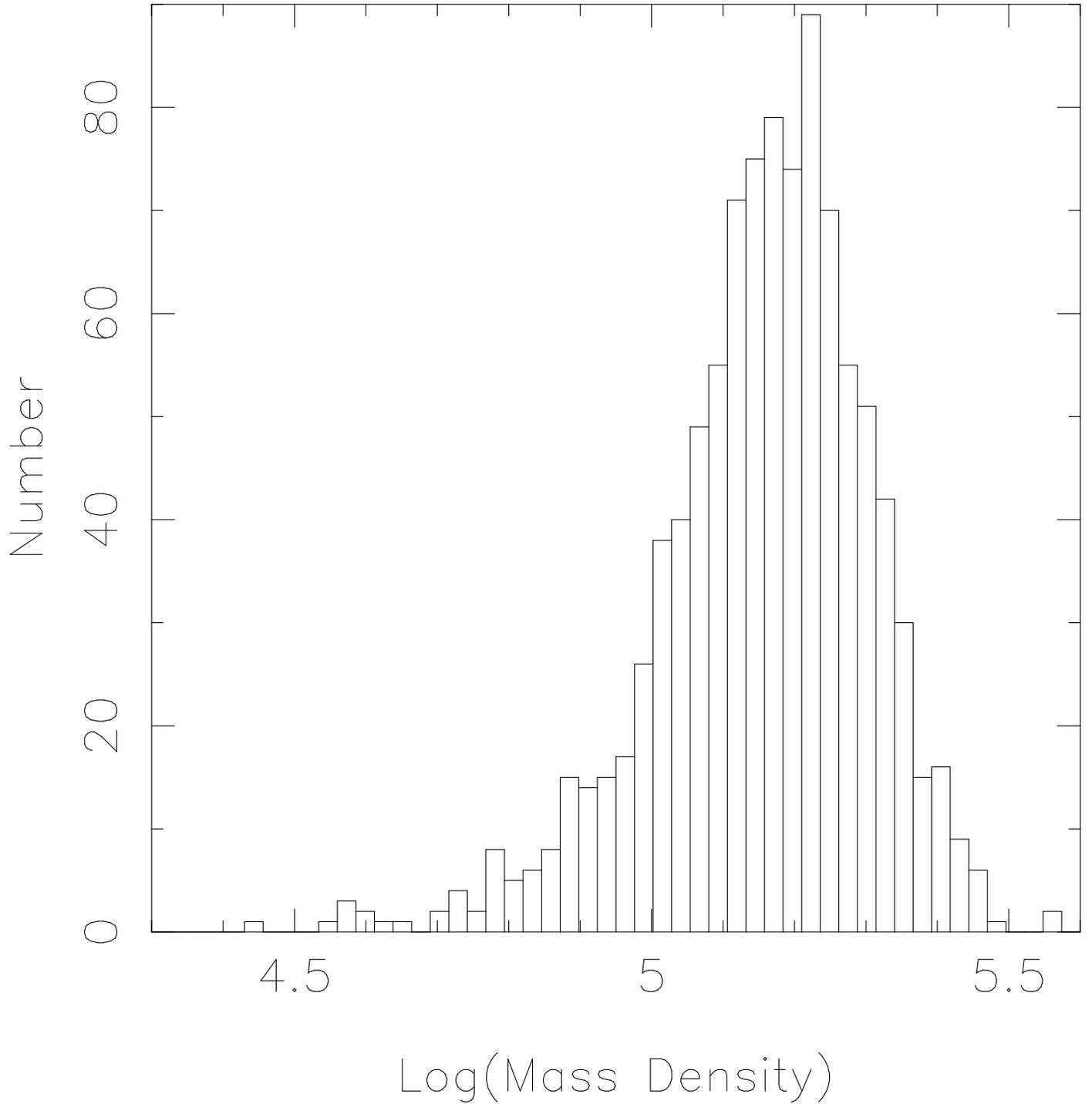}
\caption{Histogram showing the
1000 computed mass densities for
elliptical galaxies (in 
${M_\odot}{Mpc^{-3}}$) obtained using a Monte
Carlo analysis with a minimum black hole mass limit of ${10^6} M_{\odot}$.
Both the average and median densities are $1.5 \cdot 10^5 {M_{\odot}} {Mpc^{-3}}$.
\label{fig1}}
\end{figure}

\begin{figure}
\plotone{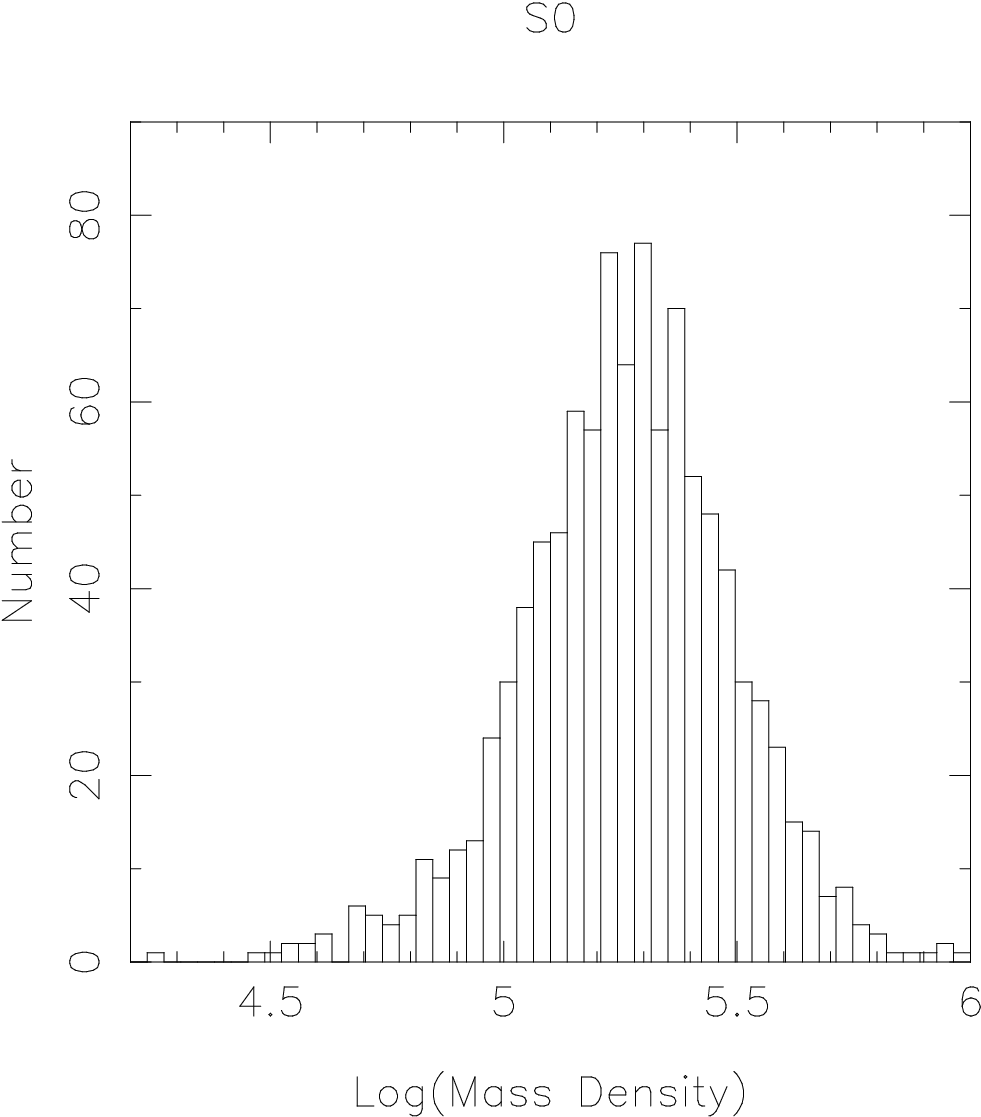}
\caption{Histogram showing the
1000 computed mass densities for S0 galaxies (in 
${M_\odot}{Mpc^{-3}}$) obtained using a Monte
Carlo analysis with a minimum black hole mass limit of ${10^6} M_{\odot}$.
The average density is 
$2.1 \cdot 10^5 {M_{\odot}} {Mpc^{-3}}$, while the
median density is $1.9 \cdot 10^5 {M_{\odot}} {Mpc^{-3}}$.
\label{fig2}}
\end{figure}

\begin{figure}
\plotone{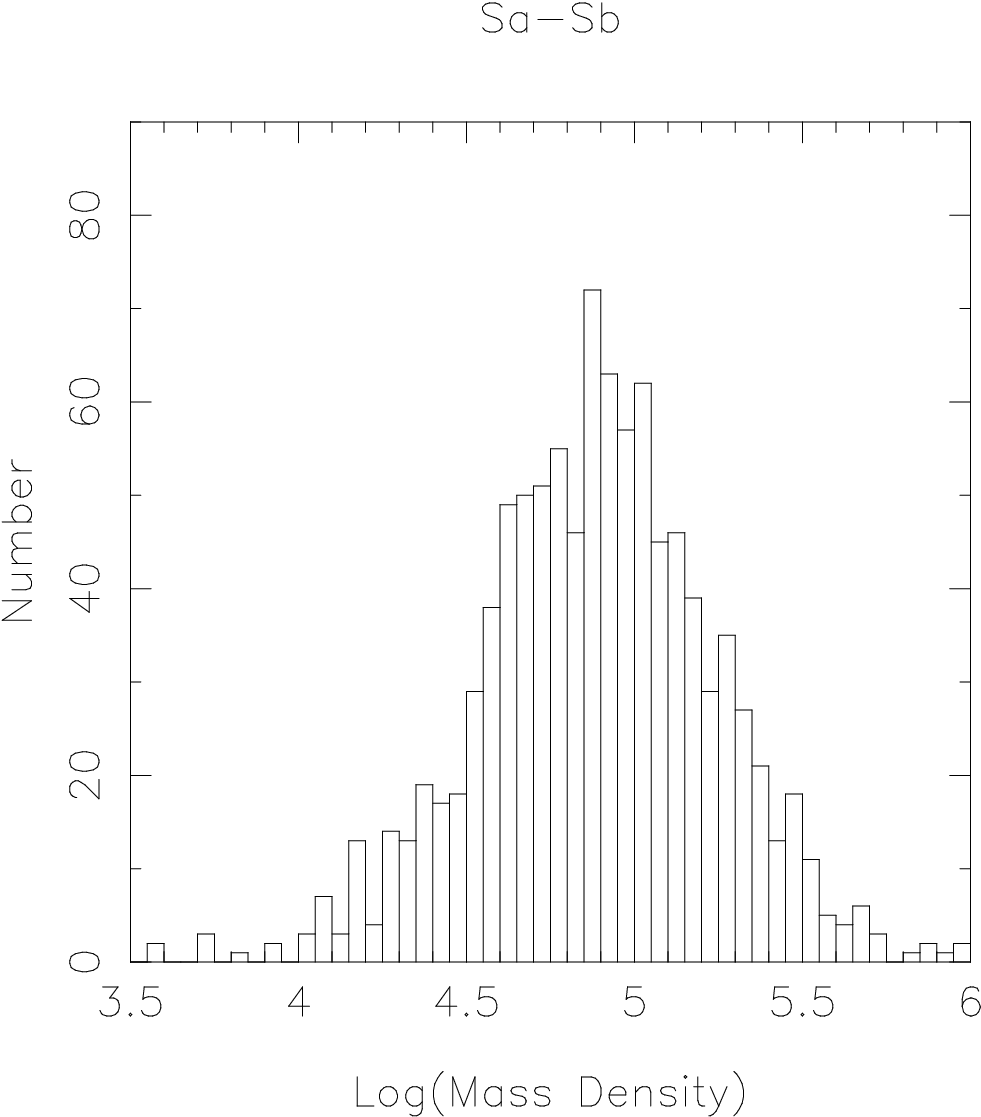}
\caption{Histogram showing the
1000 computed mass densities for Sa-Sb galaxies (in 
${M_\odot}{Mpc^{-3}}$) obtained using a Monte
Carlo analysis with a minimum black hole mass limit of ${10^6} M_{\odot}$.
The average density is 
$1.1 \cdot 10^5 {M_{\odot}} {Mpc^{-3}}$, while the
median density is $7.8 \cdot 10^4 {M_{\odot}} {Mpc^{-3}}$.
\label{fig3}}
\end{figure}
 
\begin{figure}
\plotone{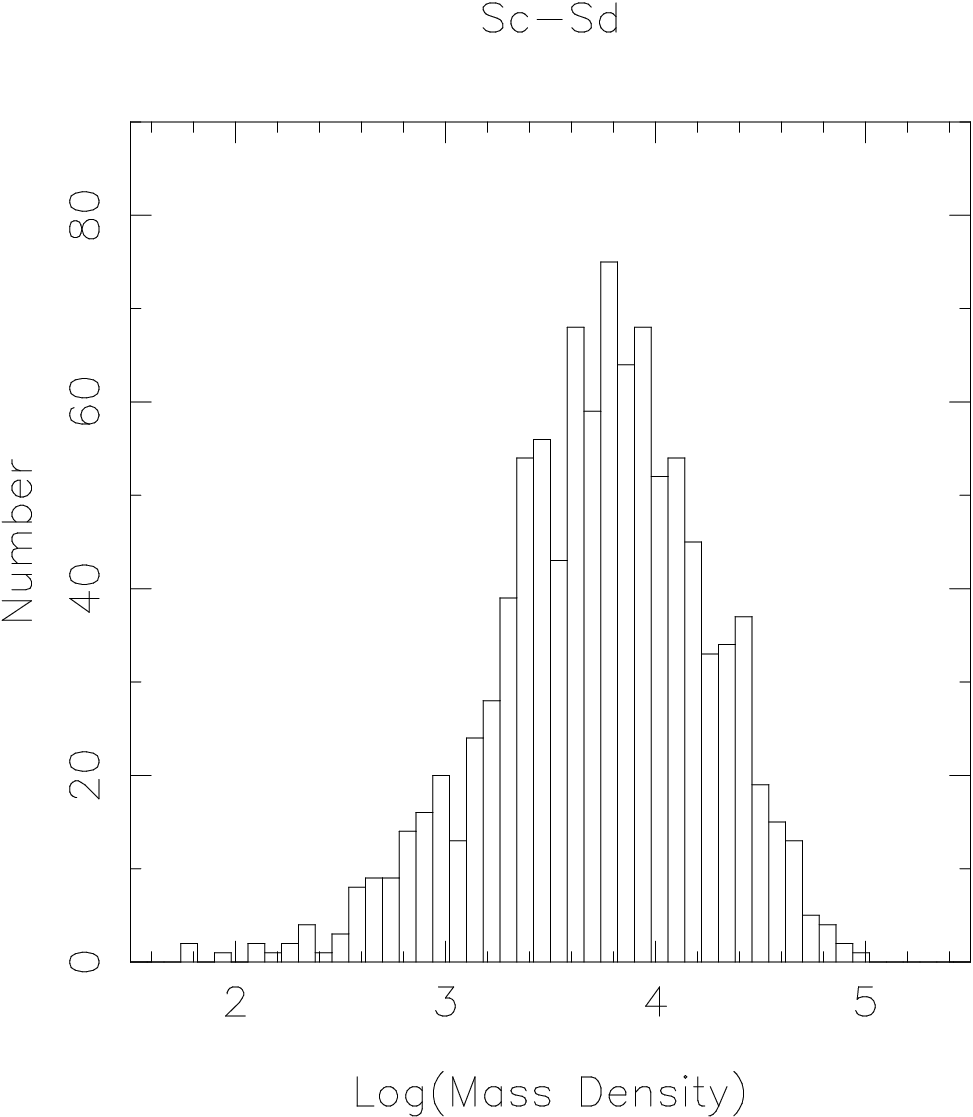}
\caption{Histogram showing the
1000 computed mass densities for Sc-Sd galaxies(in 
${M_\odot}{Mpc^{-3}}$) obtained using a Monte
Carlo analysis with a minimum black hole mass limit of ${10^6} M_{\odot}$.
The average density is     
$1 \cdot 10^4 {M_{\odot}} {Mpc^{-3}}$, while the
median density is $6 \cdot 10^3 {M_{\odot}} {Mpc^{-3}}$. 
\label{fig4}}
\end{figure}

\begin{figure}
\plotone{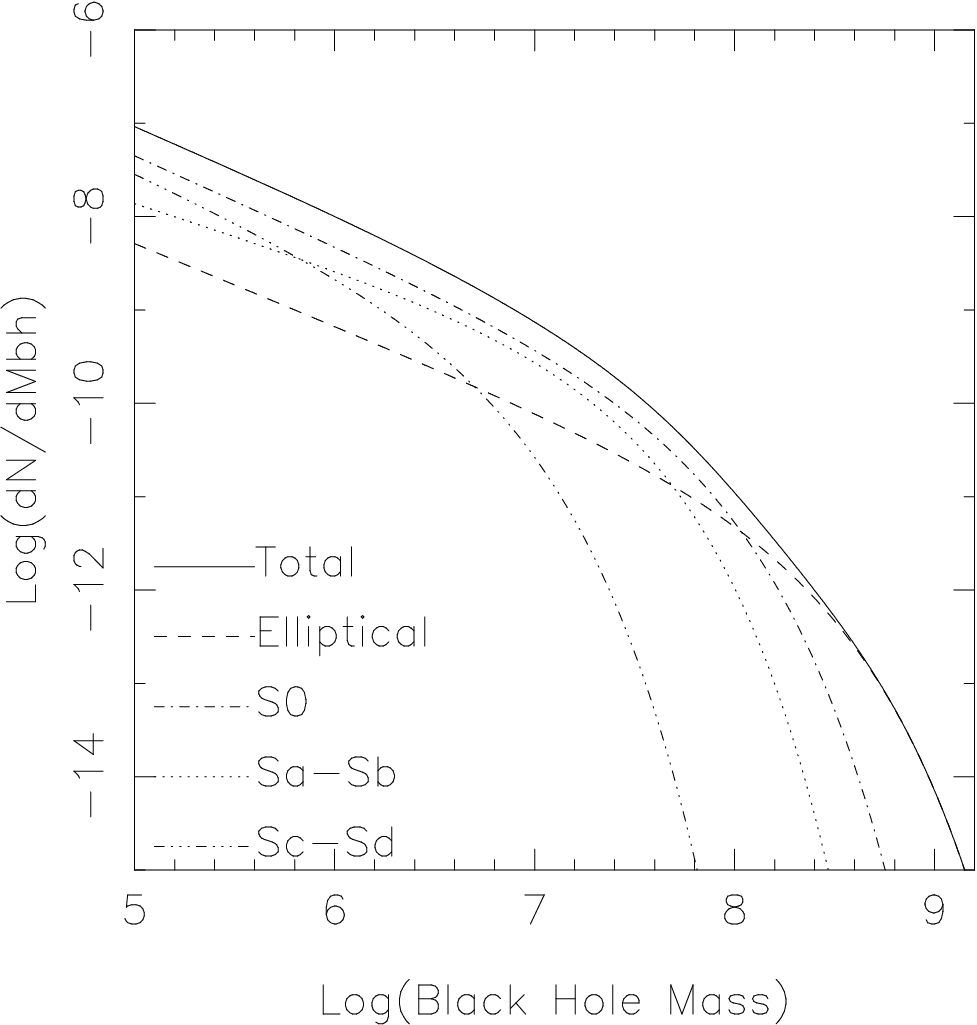}
\caption{The log of dN/$d{\cal{M}}_{\bullet}$
in ${M_\odot}^{-1}Mpc^{-3}$ as a function of the black hole
mass, ${\cal{M}}_{\bullet}$ (in $M_\odot$)for each galaxy type.
There is no information for ${\cal{M}}_{\bullet} < 10^6 M_\odot$;
the plot extends to $10^5 M_\odot$ only to illustrate the extrapolation.
\label{fig5}}
\end{figure}
 
\begin{figure}
\plotone{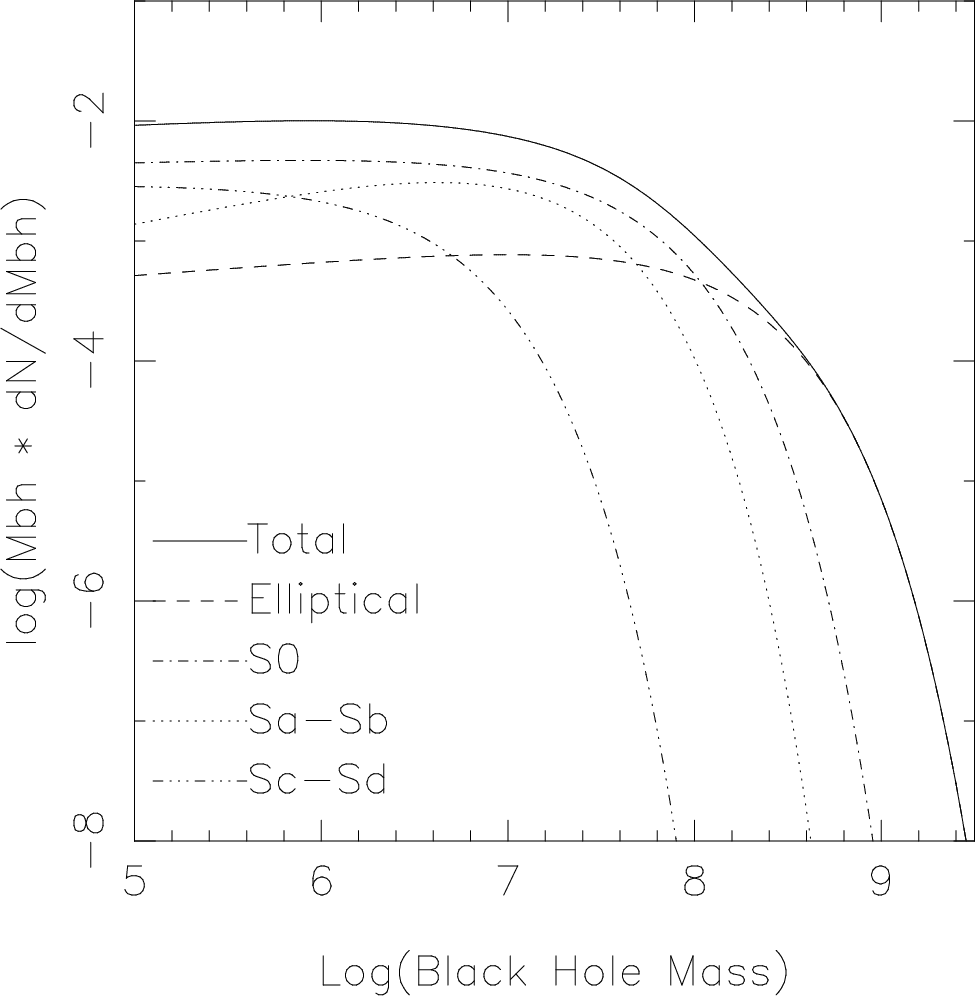}
\caption{The log of ${\cal{M}}_{\bullet}$
dN/$d{\cal{M}}_{\bullet}$ in $Mpc^{-3}$ as a function of
the black hole mass, ${\cal{M}}_{\bullet}$ (in $M_\odot$) for each galaxy type.
There is no information for ${\cal{M}}_{\bullet} < 10^6 M_\odot$; 
the plot extends to $10^5 M_\odot$ only to illustrate the extrapolation.
\label{fig6}}
\end{figure}

\begin{figure}
\plotone{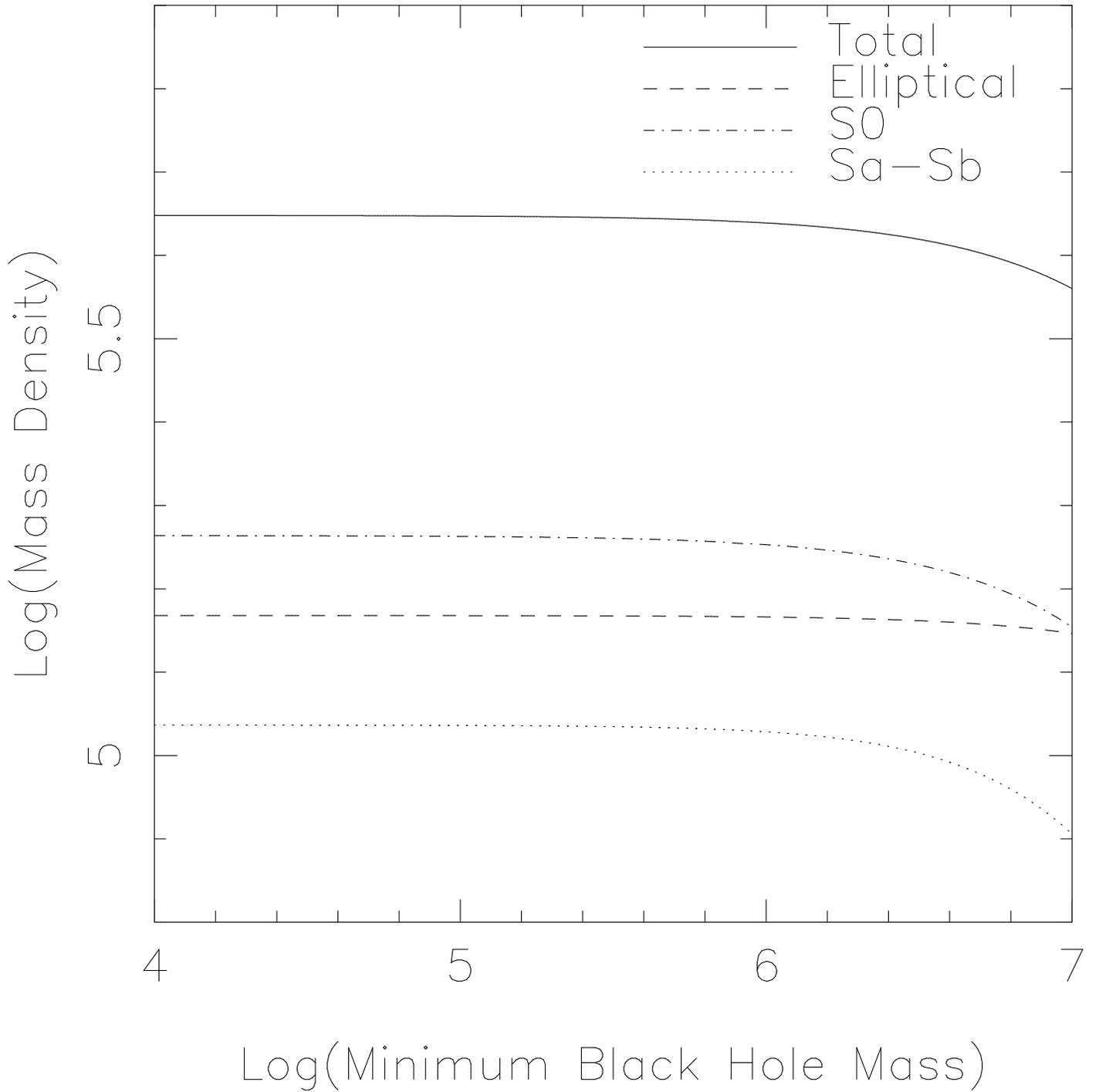}
\caption{The log of the calculated black hole
mass density (in ${M_\odot}{Mpc^{-3}}$) as a function of the minimum
black hole mass (in $M_\odot$) used in the integration of equation (27).
\label{fig7}}
\end{figure}

\begin{figure}
\plotone{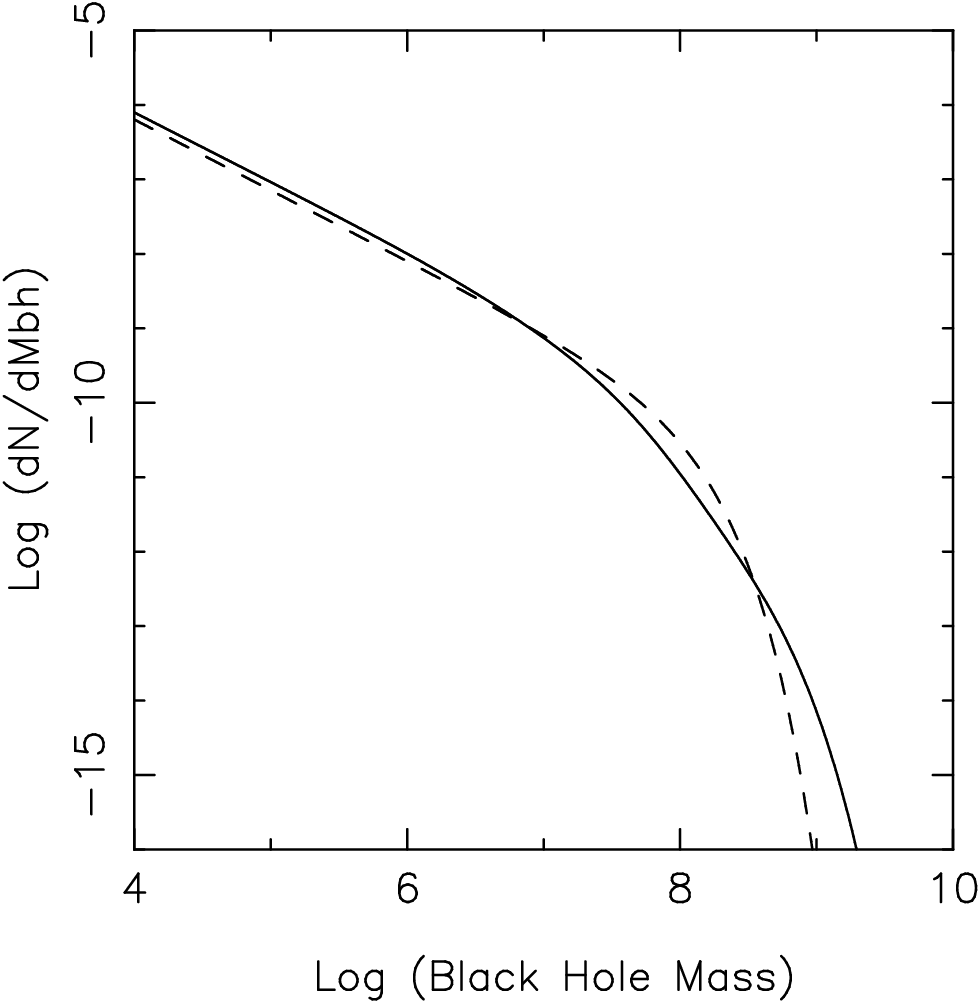}
\caption{The fit
to the dN/$d{\cal{M}}_{\bullet}$ (in ${M_\odot}^{-1}Mpc^{-3}$) 
data derived from the original \cite{Marzke} luminosity functions,
with ${\cal{M}}_{\bullet}$ in $M_\odot$.
The data is depicted by the solid line, while the dashed line is the function
fit, as given in equation 31, with fitting parameters
(c, $\cal{M}_{\bullet*}$, $\alpha$) $= (1.3 \cdot 10^{-10} {M_\odot}^{-1}Mpc^{-3},
7.9 \cdot 10^7 M_\odot, 0.95)$. 
\label{fig8}}
\end{figure}

\begin{figure}
\plotone{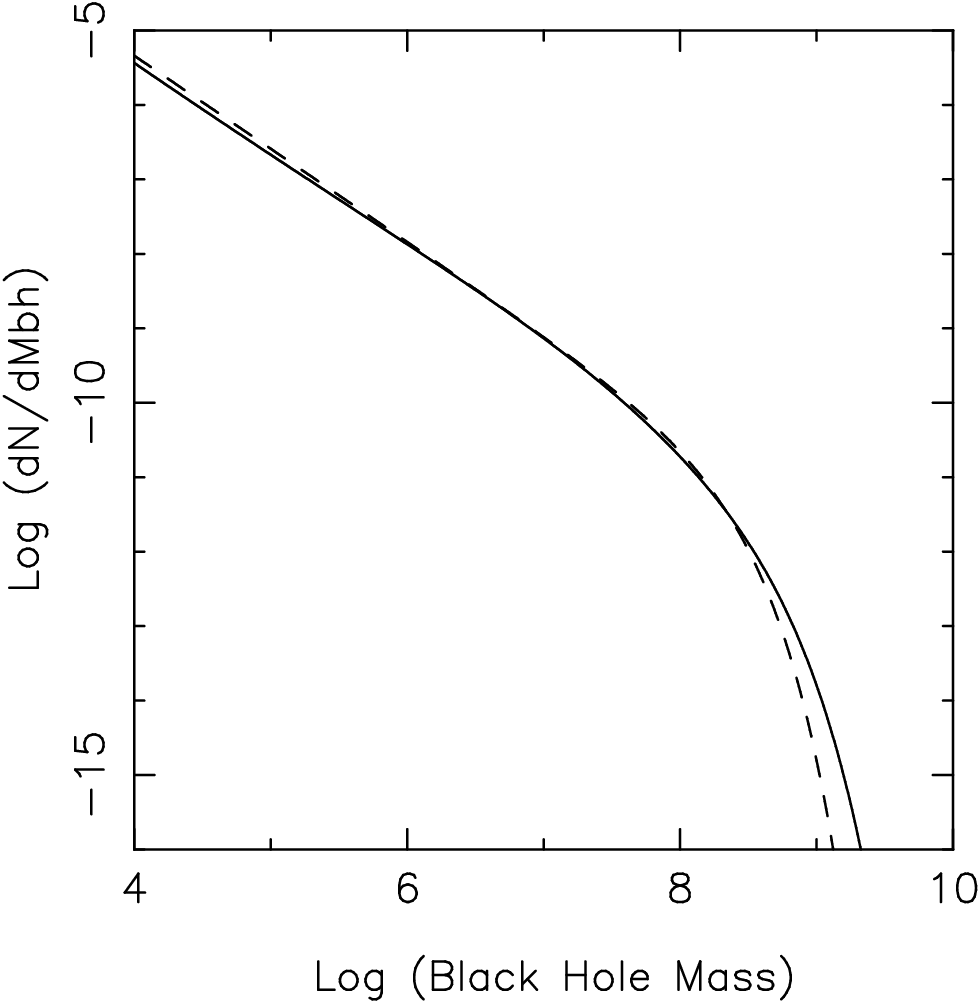}
\caption{The fit
to the dN/$d{\cal{M}}_{\bullet}$ (in ${M_\odot}^{-1}Mpc^{-3}$) 
data derived from the original \cite{Madgwick} luminosity functions
with ${\cal{M}}_{\bullet}$ in $M_\odot$.
The data is depicted by the solid line, while the dashed line is the function
fit, as given in equation 31, with fitting parameters
(c, $\cal{M}_{\bullet*}$, $\alpha$) $= (3.2 \cdot 10^{-11} {M_\odot}^{-1}Mpc^{-3},
1.3 \cdot 10^8 M_\odot, 1.25)$.
\label{fig9}}
\end{figure}

\clearpage

\begin{deluxetable}{cccrcc}
\footnotesize
\tablecaption{Fitting Parameters for Galaxies.\label{tbl-1}}
\tablewidth{0pt}
\tablehead{\colhead{Galaxy Type} & \colhead{${M_*}$} &
\colhead{${\phi_*}$} & 
\colhead{$\alpha$} & \colhead{T Types} & 
\colhead{${\Delta}$M} \\ \colhead{} & \colhead{} & \colhead{(${10^{-3}}{Mpc^{-3}}$)} &
\colhead{} & \colhead{} & \colhead{} 
}
\startdata
Elliptical & -19.23 & 1.5 $\pm$ 0.4 & -0.85 & N/A & 0. \\
S0         & -18.74 & 7.6 $\pm$ 2.0 & -0.94 & -3 - 0  & $0.64 \pm 0.30$  \\
Sa-Sb      & -18.72 & 8.7 $\pm$ 2.2 & -0.58 &  1 - 4  & $1.46 \pm 0.56$ \\
Sc-Sd      & -18.81 & 4.4 $\pm$ 1.1 & -0.96 &  5 - 7  & $2.86 \pm 0.59$ \\
\enddata

\end{deluxetable}

\clearpage

\begin{deluxetable}{ccccc}
\footnotesize
\tablecaption{Computed Mass Density for ${\cal{M}}_{{\bullet}min}= {10^6} M_{\odot}$ using \cite{Marzke}
Parameters.
\label{tbl-2}}
\tablewidth{0pt}
\tablehead{\colhead{Galaxy Type}  & 
\colhead{${\cal{M}}_{\bullet*}$} & 
\colhead{$\theta_*$} & \colhead{$\gamma$} &
\colhead {$\overline{\rho}$} \\
\colhead{} & \colhead{($10^7$$M_\odot$)} & \colhead{($10^{-3}Mpc^{-3}$)} &
\colhead{} & \colhead{($ {10^5}{M_{\odot}} {Mpc^{-3}} $)}}   
\startdata
Elliptical & $11.2 \pm 1.9$ & $1.2 \pm 0.3$ & 
$0.12 \pm 0.01$ & 
$1.5 \pm 0.5$\\ 
S0         & $3.1 \pm 1.3$ & $5.9 \pm 1.6$ & 
$0.046 \pm 0.006$ & 
$2.1 \pm 1.1$ \\ 
Sa-Sb      & $1.3 \pm 1.0$ & $6.7 \pm 1.8$ &
$0.32 \pm 0.04$ & 
$1.1 \pm 1.0$ \\ 
Sc-Sd       & $0.29 \pm 0.26$ & $3.4 \pm 0.9$ &
$0.031 \pm 0.004$ & 
$0.1 \pm 0.1$ \\ 
\enddata

\tablecomments{ For all galaxy types $\epsilon = n/\lambda = 0.8 \pm 0.1$.}

\end{deluxetable}

\clearpage
 
\begin{deluxetable}{ccccc}
\footnotesize
\tablecaption{Computed Mass Density for ${\cal{M}}_{{\bullet}min}= {10^6} M_{\odot}$
using \citet{Madgwick} Parameters. \label{tbl-3}}
\tablewidth{0pt}
\tablehead{\colhead{Madgwick Galaxy Type}  & 
\colhead{Assumed T-Type(s)} &
\colhead{${\Delta}$M} &
\colhead{${\cal{M}}_{\bullet*}$} &
\colhead {$\overline{\rho}$} \\
\colhead{} & \colhead{} & \colhead{} & \colhead{($10^7 M_{\odot}$)} &
\colhead{(${10^5}{M_{\odot}} {Mpc^{-3}} $)}
}
\startdata
Type 1 ($\eta < -1.4$) & Elliptical & 0. & $10.0 \pm 1.8$ & $3.0 \pm 0.6$ \\
Type 1 ($\eta < -1.4$) & -3 - 0 & $0.64 \pm 0.30$ & $4.9 \pm 2.0$ & $1.5 \pm 0.6$ \\ 
Type 1 ($\eta < -1.4$) & 1 - 2 & $1.19 \pm 0.51$ & $2.7 \pm 1.7$ & $0.81 \pm 0.50$ \\
Type 2 ($-1.4 \leq \eta < 1.1$) & 1 - 5 & $1.58 \pm 0.60$ & $1.8 \pm 1.3$ & $1.1 \pm 0.9$ \\
Type 3 ($1.1 \leq \eta < 3.5$) & 3 - 5 & $1.85 \pm 0.59$ & $0.86 \pm 0.70$ & $ 0.37 \pm 0.36$ \\ 
Type 4 ($\eta \ge 3.5$) & 5 & $2.47 \pm 0.45$ & $0.37 \pm 0.26$ & $0.07 \pm 0.07$ \\ 
\enddata
 
\end{deluxetable}


\begin{thebibliography}{}

\bibitem[Chokshi \& Turner (1992)]{Chokshi}
Chokshi, A. \& Turner, E.L. 1992, \mnras, 259, 421  

\bibitem[de Vaucouleurs \& Olson (1982)]{deV}
de Vaucouleurs, G., \& Olson,  D.W. 1982, \apj, 256, 346

\bibitem[Faber \& Jackson (1976)]{fj}
Faber, S.M., \& Jackson, R.E. 1976, \apj, 204, 668

\bibitem[Fabian \& Iwasawa (1999)]{Fabian}
Fabian, A.C. \& Iwasawa, K. 1999, \mnras, 303, L34 

\bibitem[Ferrarese \& Merritt (2000)]{Merritt}
Ferrarese, L. \& Merritt, D. 2000, \apj, 539, L9

\bibitem[Gebhardt et al. (2000)]{Gebhardt}
Gebhardt, K., et al. 2000, \apj, 539, L13 

\bibitem[Gonzalez et al. (2000)]{Gonzalez}
Gonzalez, A.H., Williams, K.A., Bullock, J.S., Kolatt, T.S., \& Primack, J.R. 
2000, \apj, 528, 145

\bibitem[Kormendy \& Gebhardt (2002)]{Kormendy}
Kormendy J. \& Gebhardt, K. 2002, in 20th Texas Symposium on Relativistic
Astrophysics, ed. H. Martel \& J.C. Wheeler, AIP (astro-ph/0105230)

\bibitem[Kormendy \& Richstone (1995)]{KR}
Kormendy, J. \& Richstone, D. 1995, \araa, 33, 581 

\bibitem[Loveday et al. (1992)]{Loveday}
Loveday, J., Peterson, B.A., Efstathiou, G., \& Maddox, S.J., 1992, \apj,
390, 338

\bibitem[Madgwick et al. (2002)]{Madgwick}
Madgwick, D.S., et al. 2002, \mnras, in press (astro-ph/0107197) 

\bibitem[Magorrian (1998)]{Magorrian}
Magorrian, J. et al. 1998, \aj, 115, 2285

\bibitem[Marzke et al. (1994)]{Marzke}
Marzke, R.O., Geller, M.J., Huchra, J.P., \& Corwin, H.G. 1994, \aj, 108, 437

\bibitem[Marzke, Huchra, \& Geller (1994)]{MHG}
Marzke, R.O., Huchra, J.P., \& Geller, M.J. 1994, \apj, 428, 43

\bibitem[Merritt \& Ferrarese (2001a)]{MF}
Merritt, D., \& Ferrarese, L. 2001a, \mnras, 30, 320L

\bibitem[Merritt \& Ferrarese (2001b)]{Merritt2}
Merritt, D., \& Ferrarese, L. 2001b, in The Central kpc of
Starbursts and AGN, eds. J.H. Knapen et al. (San Francisco:
Astronomical Society of the Pacific), in press (astro-ph/0107134) 

\bibitem[Richstone et al. (1998)]{Richstone}
Richstone, D. et al. 1998, \nat, 395, A14 

\bibitem[Salucci et al. (1999)]{Salucci}
Salucci, P., Szuszkiewicz, E., Monaco, P., \& Danese, L. 1999, \mnras, 307, 637

\bibitem[Shanks et al. (1984)]{Shanks}
Shanks, T., Stevenson, P.R.F., Fong, R., \& MacGillivray, H.T. 1984, \mnras, 206, 767

\bibitem[Simien \& de Vaucouleurs (1986)]{Simien}
Simien, F., \& de Vaucouleurs, G. 1986, \apj, 302, 564

\bibitem[Soltan (1982)]{Soltan}
Soltan, A. 1982, \mnras, 200, 115

\bibitem[Tremaine et al. (2002)]{Tremaine}
Tremaine, S. et al. 2002, in press (astro-ph/0203468) 

\bibitem[Yu \& Tremaine (2002)]{Yu}
Yu, Q. \& Tremaine, S. 2002, \mnras, in press (astro-ph/0203082)

\end{thebibliography}
\end{document}